
%
\input phyzzx
\sequentialequations

\catcode`\@=11
\def\PLsection#1{\par \penalty-300 \vskip\chapterskip
   \spacecheck\chapterminspace
   \noindent {\twelvebf #1}\par
   \nobreak\vskip\headskip \penalty 30000 }

\def\refitem#1{\r@fitem {[#1]} }
\def\figend@{\rel@x {\number\figurecount}}
\def\Fignum#1{\FIGNUM #1\figend@ }
\def\Fig#1{\Fignum #1\FIGWRITE }
\def\figitem#1{\r@fitem{Figure #1.}}
\catcode`\@=12

\def\simge{\rlap{\raise 2pt \hbox{$>$}}{\lower 2pt \hbox{$\sim$}}}
\def\simle{\rlap{\raise 2pt \hbox{$<$}}{\lower 2pt \hbox{$\sim$}}}

\def\sec{ \,{\rm s} } \def\persec{ \,{\rm s}^{-1} } \def\yr{ \,{\rm yr} }
 \def\eV{ \,{\rm eV} } \def\MeV{ \,{\rm MeV} }
\def\perMeV{ \,{\rm MeV} } \def\GeV{ \,{\rm GeV} } \def\perGeV{ \,{\rm
GeV}^{-1} } \def\TeV{ \,{\rm TeV} }   \def\sqcm{ \,{\rm cm}^2 } \def\sqkm{ \,{\rm km}^2 } \def\persqcm{
\,{\rm cm}^{-2} } \def\percc{ \,{\rm cm}^{-3} } 
\def\persr{ \,{\rm sr}^{-1} } 
\def\persqcmssrMeV{ \persqcm \persec \persr \perMeV } \def\persqcmssrGeV{
\persqcm \persec \persr \perGeV }

\nopubblock
\vbox to 1in {
\hbox to 6.5in {\hfill Uppsala U.\ PT17-1992}
\hbox to 6.5in {\hfill June 1992}
\vss }
\titlepage
\title{\fourteenbf Indirect detection of unstable heavy dark matter}
\author{Paolo Gondolo\footnote{1}{BITNET address: GONDOLO@SESUF51}}
\address{Department of Radiation
Sciences, Uppsala University, \break P.O.\ Box~535, 75121 Uppsala, Sweden}
\vskip 1cm

\abstract
Unstable relics with lifetime longer than the age of the Universe could be the
dark matter today. Electrons, photons and neutrinos are a natural outcome of
their decay and could be searched for in cosmic rays and in $\gamma$-ray and
neutrino detectors. I compare the sensitivities of these three
types  of searches to the mass and lifetime of a generic unstable particle. I
show  that if the relics constitute our galactic  halo and their branching
ratios into electron-positrons, photons and neutrinos are comparable, neutrino
searches would probe the longest lifetimes  for masses $\simge 40 \TeV$, while
electron-positron searches would be better but more uncertain for lighter
particles. If instead the relics are not clustered in  our halo, neutrinos are
more sensitive a probe than $\gamma$-rays for masses $\simge 700 \GeV$.  A $ 1
\sqkm $ neutrino telescope should be able to explore lifetimes up to  $ \sim
10^{30} \sec $ while searching for neutrinos from unstable particles  above the
atmospheric background.
\endpage

\PLsection{1.~~Introduction}

Popular elementary particle candidates for cold dark matter (e.g.\ neutrinos,
cosmions and sneutrinos) have been ruled out or severely constrained by the SLC
and LEP measurements of the $Z$ decay width combined with the non-observation
of a signal in direct and indirect dark matter searches, in low-background
germanium detectors and proton-decay experiments respectively (see e.g.\
ref.~[\Ref\GriestSilkKrauss{ K.~Griest and J.~Silk, Nature 343 (1990) 26;
\nextline L.~Krauss, Phys.\ Rev.\ Lett.\ 64 (1990) 999.}]).

It is so natural that alternative possibilities for non-baryonic cold dark
matter are being (re)explored. Here I consider a class of unstable dark matter
candidates that are heavy and long-lived, decaying on cosmological time scales.
Many particles of this sort have been  already proposed before the
above-mentioned experimental results for various reasons.

Massive long-lived particles with a substantial relic density may arise in
technicolor models, which are an interesting alternative to the standard Higgs
mechanism for spontaneous $SU(2) \times U(1)$ symmetry breaking. It was pointed
out in ref.~[\Ref\Nussinov{ S.~Nussinov, Phys.\ Lett.\ B165 (1985) 55.}] that
the lightest technibaryon, which likely has mass $ m \sim \TeV $ and  lifetime
$10^{27-32} \sec $ could account for the missing mass if there is a
technibaryon-antitechnibaryon asymmetry comparable to the baryon-antibaryon
asymmetry. Non-perturbative fermion-number violating processes in the Standard
Model could generate such an asymmetry in a natural way [\Ref\BarrChivikula{
S.M.~Barr, R.S.~Chivikula and E.~Farhi, Phys.\ Lett.\ B241 (1990) 387.}].

Motivated by a preliminary report of an anomalous high energy ($>$ 10 GeV)
positron component in cosmic rays which is unlikely to be generated by
spallation processes, new massive long-lived particles have been proposed as
dark matter candidates. It has been suggested that the excess positrons come
from the decay of $\sim$ 30 GeV right-handed neutrinos with lifetime $\sim
10^{25} \sec$ [\Ref\BabuEichlerMohapatra{ K.S.~Babu, D.~Eichler and
R.N.~Mohapatra, Phys.\ Lett.\ B226  (1989) 347.}] or from 1-3 TeV GUT particles
with lifetime $\sim 10^{24} \sec$ [\Ref\Eichler{ D.~Eichler, Phys.\ Rev.\
Lett.\ 63 (1989) 2440.}].

The `lightest supersymmetric particle' can be unstable and long-lived in
supersymmetric models with very weak R-parity breaking. A decaying gravitino of
mass $ \sim 100 \GeV $  [\Ref\Berezinsky{ V.~Berezinsky, Bartol preprint
BA-90-87 (1990); Gran Sasso preprint LNGS-91-02 (1991).}] and a
neutralino-neutrino mixed state of mass $ 10 - 50 \GeV $
[\Ref\BerezinskyMasieroValle{ V.~Berezinsky, A.~Masiero and J.W.F.~Valle,
Phys.\ Lett.\ B266 (1991) 382.}], both with any lifetime longer than $ \sim
10^{17} \sec $, have been examined in a cosmological context.

As a final example, a simple solution via confinement [\Ref\Antoniadis{
I.~Antoniadis, J.~Ellis, J.~Hagelin and D.V.~Nanopoulos, Phys.\ Lett.\ B231
(1989) 65.}] to the problem of fractional charges in models derived from the
superstring [\Ref\Schellekens{ A.N.~Schellekens, Phys.\ Lett.\ B237 (1990)
363.}] results in  several integer-charged composite particles, named
`cryptons'
[\Ref\cryptons{J.~Ellis, J.~Lopez and D.V.~Nanopoulos, Phys.\ Lett.\ B247
(1990) 257.}], some of which could be long-lived with lifetimes as long as $
10^{23} \sec$ and masses $\sim 10^{12} \GeV $ and could, in principle,
constitute the dark matter [\Ref\EGLNS{ J.~Ellis, G.B.~Gelmini, J.~Lopez,
D.V.~Nanopoulos and S.~Sarkar, Nucl.\ Phys.\ B373 (1992) 399.}]. Their expected
relic density is however very uncertain since it depends on the amount of
entropy released in the decays of short-lived cryptons (and other particles)
after the long-lived cryptons go out of chemical equilibrium.

It is therefore interesting to examine long-lived heavy particles, with masses
$\simge \GeV$ and lifetimes $\simge 10^{17} \sec$, as dark matter candidates.
At present, it seems appropriate to study such weakly unstable massive
particles (WUMPs)\footnote{*}{I adopt the terminology of
ref.~[\BabuEichlerMohapatra].} in a model-independent way.

In general, a WUMP $x$ is characterized by its mass $m_x$, its lifetime
$\tau_x$ and its branching fractions $b_y$ into different decay channels $y$. I
focus here on WUMPs which constitute the dark matter today and I consider two
classes of WUMP distributions: a uniform mass distribution throughout the
Universe with critical value $\rho_c = 10.5 \,{\rm keV} \percc h^2 $
(unclustered WUMPs) and an inhomogeneous distribution with WUMPs clustered in
our galactic halo with mass density $\rho_\odot = 0.3 \GeV \percc$ in the solar
neighborhood (clustered WUMPs). Other values of the relic WUMP density have
been considered in refs.~[\EGLNS,\Ref\GGS{P.~Gondolo, G.B.~Gelmini and
S.~Sarkar, UCLA preprint UCLA/91/TEP31.}]. As usual, $h$ is the Hubble constant
in units of 100 km s${}^{-1}$ Mpc${}^{-1}$.

Of the WUMP decay products, I consider here neutrinos, electrons and photons,
which may give origin to diffuse extraterrestrial fluxes.  In the following, I
examine the range of WUMP masses and lifetimes that could be explored by
background- and flux-limited detectors searching for such indirect signals from
unstable dark matter particles.

\PLsection{2.~~Diffuse photon, electron-positron and neutrino fluxes}

Branching ratios and energies of the decay products, and even the nature of the
decay products, depend of course on the particular model for the decaying dark
matter.  For the sake of definiteness, I imagine that WUMPs undergo 3-body
decays, in which one (or more) decay product(s) is a photon, an electron, a
positron or a neutrino, and acquires a typical energy $E_0 = m_x / 3 $.
Furthermore, when comparing sensitivities to signals of different origin, I
take all branching ratios $b_y$ to be equal, in particular I use $b_y = 1$ when
stating limits on the lifetime $\tau_x$. All results can be trivially rescaled
to the values appropriate to a specific model. More complicated generation
scenarios, e.g.\ secondary products or jets, can also be easily incorporated by
including the decay product multiplicity and possible continuous energy
spectra in an energy dependent $b_y$.

The maximum WUMP lifetime explorable by background-limited detectors is obtain
as a function of the WUMP mass by imposing that the electron-positron, photon
and neutrino fluxes from WUMP decays be smaller than the respective
backgrounds.  I separately discuss now the three cases of decay photons,
electron-positrons and neutrinos. I will use $t_0=10^{10}\yr$ and
$h=1$.

\PLsection{2.1~~Photons}

The photon flux expected from uniformly distributed decaying dark matter is a
superposition of the photon spectra generated in decays occuring at different
times. The resulting present day flux $I_\gamma(E)$ can be written as an
integral over the redshift $1+z = E'/E$ of the photon spectrum per decaying
particle $S_\gamma(E)$: $$ I_\gamma(E) = { 3 \over 8 \pi} { \rho_c t_0 \over
m_x \tau_x } E^{1/2} \int_E^\infty { {\rm d} E' \over E'^{3/2} } S_\gamma(E') .
\eqn \ugaflux $$ (This equation is obtained for a matter-dominated universe and
for $\tau_x \gg t_0$.)

For WUMPs clustered in the galactic halo no integration over redshift is
necessary and the decay photon flux is given by $$  I_\gamma(E) = { 1 \over 4
\pi} \, { 1 \over m_x \tau_x} \, \int { \rho_{\rm h}({\bf x}) \over | {\bf x} -
{\bf x}_\odot |^2 } d^3 x \, S_\gamma(E) , \eqn \cgaflux $$ where ${\bf
x}_\odot$ denotes the position  of the solar system and $\rho_{\rm h}({\bf x})$
is the WUMP distribution in the halo. If the dark matter
distribution is taken to be $$ \rho_{\rm h} ({\bf x}) = { 2 \rho_\odot \over 1
+ \left( r / a \right) ^2 }, \eqn \grho $$ the integral in eq.~\cgaflux\
evaluates to $ 31.0 a \rho_\odot $ and for $\rho_\odot = 0.3 \GeV \percc $ and
$ a_\odot = 8 \, \rm{kpc} $, it is $\simeq  2.3 \rho_c t_0 $.

Two cases arise for the source function $S_\gamma(E)$ according to the energy
of   the primary decay photon $E_0$. If it is energetic enough to produce
$e^+e^-$ pairs in collisions against the microwave photons, it can  trigger
electromagnetic cascades. Otherwise it is able to propagate to us without
cascading. In the latter case, the source spectrum can be approximated by a
line $$ S_\gamma(E) \simeq b_\gamma \, \delta(E-E_0) . \eqn \nocascade $$

The cascading case is more complicated. The electromagnetic cascade developes
until the energies of the photons have fallen below the pair production
threshold $E_{\rm max}$.  For a blackbody target this is given by
[\Ref\ZdiarskiSvensson{ A.~Zdiarski and R.~Svensson, Astrophys.\ J.\ 344 (1989)
551.}] $$ E_{\rm max} \simeq { m_e^2 \over 20.4 T \left[ 1 + {1\over2}
\ln\left({\eta\over7\times10^{-10}}\right)^2 + {1\over2} \ln\left({E_{\rm
max}\over m_e} \right)^2 \right] } , \eqn \ecrit $$ where $m_e$ is the electron
mass, $T$ is the blackbody temperature and $\eta$ is the baryon to photon
ratio, which for simplicity I take to be just $7\times10^{-10}$. At the present
epoch, $ E_{\rm max}(t_0) \simeq 3.4 \times 10^{12} \eV$ and
cascades can be generated if the WUMP mass $m_x \simge m_{\rm crit}
\simeq 10^{13} \eV$. The spectrum of the `breakout'
photons below the pair-production threshold falls as $ \sim E^{-1.5}$ until
$\sim 0.04 E_{\rm max} $ and then steepens to $ \sim E^{-1.8} $ before being
cutoff at $E_{\rm max}$ [\Ref\Zdiarski{ A.~Zdiarski, Astrophys.\ J.\ 335 (1988)
786.}]. With the normalization $ \int {\rm d} E E S_\gamma(E) = E_0 $, the
source spectrum per decaying WUMP with $ m_x \simge 3 E_{\rm max} $ is $$
S_\gamma(E) \simeq \cases{ {3\over4} b_\gamma E_0 E_{\rm max}^{-1/2} E^{-3/2},
&
for $ 0 \le E \le 0.04 E_{\rm max}$, \cr {3\over10} b_\gamma E_0 E_{\rm
max}^{-0.2} E^{-1.8}, & for $ 0.04 E_{\rm max} \le E \le E_{\rm max}, $ \cr 0,
&
for $ E \ge E_{\rm max} $. \cr } \eqn \cascade $$

The source functions~\nocascade\ and~\cascade\ can then be inserted in
eqs.~\ugaflux\ and~\cgaflux\ and the resulting photon fluxes can be compared
with the diffuse background $\gamma$ flux, which at $E>3 \MeV$
I approximate as (see ref.~[\Ref\ResselTurner{ M.D.~Ressel and M.S.~Turner,
Comments Astrophys.\ 14 (1990) 323.}]) $$ I_\gamma^{\rm bkgd} \, \simle \, 2
\times 10^{-3} \left( 3 \MeV \over E \right)^2 \persqcmssrMeV. \eqn \gaflux $$
I assume a detector with 10\% energy resolution to smear the delta function
occuring in the decay neutrino flux from non-cascading clustered WUMPs.

The maximum WUMP lifetimes accessible to $\gamma$-ray astronomy turn then out
to be $$ \tau_x \simge \cases{ 1.73 \times 10^{25} b_\gamma  \sec, & for $ m_x
< m_{\rm crit}$, \cr 1.46 \times 10^{24} b_\gamma \sec, & for $ m_x > m_{\rm
crit}$, \cr} \eqn \ugatau $$ for unclustered WUMPs and $$ \tau_x \simge \cases{
2.6 \times 10^{26} b_\gamma \sec, & for $ m_x < m_{\rm crit}$, \cr 6.0 \times
10^{26} b_\gamma \sec, & for $ m_x > m_{\rm crit}$, \cr} \eqn \ugatau $$ for
clustered WUMPs. These lifetimes (with $b_\gamma=1$) are shown in
Fig.~\Fig\figone{Maximum accessible WUMP lifetime $\tau_x$ versus WUMP mass
$m_x$ in electron-positron, photon and neutrino background-limited detectors.
Light and heavy lines correspond to clustered and unclustered WUMP
distributions respectively. The dotted lines indicate the present lower limits
on $\tau_x$ imposed by the IMB and Fly's Eye neutrino data and the region
explorable by a neutrino telescope sensitive to 1 muon km$^{-2}$ yr$^{-1}$
(curve $10^6$).} as the solid lines labelled $\gamma$. The different directions
of the step at $m_{\rm crit} \simeq 10^{13} \eV$ are due to the rapid falling
of the background flux~\gaflux\ with energy combined with the fact that the
ratio between the WUMP-generated and the background fluxes is maximum at
$E=E_{\rm max}(t_0)$ in the clustered case and at $E\simeq 0.18E_{\rm
max}(t_0)$ in the unclustered one.

\PLsection{2.2~~Electrons and positrons}

If the decaying dark matter is clustered in the galactic halo, the galactic
magnetic field is able to contain the decay electrons and positrons in the
galaxy. The containment time is quite uncertain, typically thought to be of the
order of $10^{16}\sec$ for 1 GeV electrons, and possibly varying with energy.
This is a source of uncertainty in the determination of the electron-positron
fluxes from clustered WUMPs, and renders the corresponding bounds on the
lifetimes less
reliable than in the photon and neutrino cases.

For these reason, it is therefore sufficient to describe the high energy
$e^\pm$ density $n_e(E)$ using a leaky box model
[\Ref\Berezinskyetal{V.S.~Berezinsky, S.V.~Bulanov, V.A.~Dogiel, V.L.~Ginzburg
and V.S.~Ptuskin, Astrophysics of Cosmic Rays (Elsevier, 1990).}], with the
$e^\pm$ sources uniformly distributed over the halo, $$ { n_e(E) \over
\tau_{\rm cont}(E) } + { {\rm d} \over {\rm d} E } \left[ f(E) n_e(E) \right] =
Q_e(E). \eqn \leakybox $$ Here $f(E) = - \beta E^2 $, with $ \beta \sim 3\times
10^{-17} \perGeV \persec $, is the rate of energy loss of electrons and
positrons due to synchrotron radiation and inverse-Compton scattering off
microwave photons, $Q_e(E)$ is the $e^\pm$ source function, which I take to be
$$ Q_e(E) = b_e {\rho_\odot \over m_x \tau_x} \, \delta( E - E_0 ) , \eqn
\esource $$ and $\tau_{\rm cont}(E)$ is the containment time in the halo. This
can be estimated using a diffusion model [\Berezinskyetal] in which
$\tau_{\rm cont}(E) \sim R_{\rm halo}^2 / D(E)$, taking for the halo size $
R_{\rm halo} \sim 10 {\rm kpc} $ and for the diffusion coefficient $D(E) \sim
10^{29} E_{\rm GeV}^{1/3} \sqcm \persec$ [\Berezinskyetal]. The containment
time
results $ \tau_{\rm cont}(E) \sim 10^{16} \sec \, E_{\rm GeV}^{-1/3} $.  The
electron-positron flux is then obtained by solving eq.~\leakybox: $$
I_{e^\pm}(E) = b_e {1 \over 4\pi} {\rho_\odot \over m_x \tau_x \beta E^2}
\exp\left[ \left( { E_c \over E_0 } \right) ^{2/3} - \left( { E_c \over E }
\right) ^{2/3} \right], \eqn \ceflux $$ with $ E_c = \left[ {2\over 3} \beta
\tau_{\rm cont}(1\GeV) \right]^{-3/2} \sim 10 \GeV$.

The combined cosmic-ray electron and positron flux has been measured up to
about 2 TeV [\Ref\Nishimura{J.~Nishimura et al., Astrophys.\ J.\ 238 (1980)
394.}]. I adopt here the following parametrization for it (cfr.\
ref.~[\Ref\Tang{ K.-K.~Tang, Astrophys.\ J.\ 278 (1984) 881.}]): $$ I^{\rm
bkgd}_{e^\pm}(E) \simeq \cases{ 4.4 \times 10^{-7} \left( { E \over 20 \GeV }
\right) ^{-2.7} \persqcmssrGeV, & for $ E < 20 \GeV$, \cr 4.4 \times 10^{-7}
\left( { E \over 20 \GeV } \right) ^{-3.5} \persqcmssrGeV, & for $ E > 20
\GeV$. \cr} \eqn \eflux $$

Comparing the decay and background fluxes~\ceflux\ and~\eflux, the maximum WUMP
lifetime accessible to background-limited electron-positron searches is obtain
as $$ \tau_x \simge \cases{ 3.8 \times 10^{27} b_e m_{10}^{-0.3} \sec, & for
$m_x<60\GeV$, \cr 9.5 \times 10^{26} b_e m_{10}^{1/2} \sec, & for
$m_x>60\GeV$.\cr} \eqn \etau $$ This is shown in Fig.~\figone\ for $b_e=1$ as
the light line labelled $e^\pm$. The dashed portion corresponds to the
extrapolation of the background $e^\pm$ flux~\eflux\ above the highest measured
energies.

\PLsection{2.3~~Neutrinos}

The diffuse neutrino flux from decays of unclustered WUMPs is obtained  by
integration over redshift of a monochromatic decay spectrum with energy $E_0$.
For lifetimes $\tau_x \gg t_0$, it is given by $$ I_\nu(E) = b_\nu \, { 3
\over 8 \pi} \, { \rho_c \over m_x E_0 } \, { t_0 \over \tau_x } \, \left( { E
\over E_0 } \right) ^{1/2} . \eqn \unuflux $$ This equation applies separately
to each neutrino and antineutrino flavor.

As a background to the decay neutrino flux I consider the $\nu_\mu+\bar\nu_\mu$
atmospheric neutrinos produced in collisions of cosmic rays with nuclei of the
upper atmosphere. Their spectrum in the vertical direction has been estimated
in ref.~[\Ref\Volkova{ L.V.~Volkova, Sov.\ J.\ Nucl.\ Phys.\ 31 (1980) 784.}]
as
$$ I^{\rm bkgd}_{\nu_\mu+\bar\nu_\mu} \simle \left( 4.4 \, E_{\rm GeV}^{-3.69}
+ 2.4 \times 10^{-5} \, E_{\rm GeV}^{-2.65} \right) \persqcmssrGeV, \eqn
\nufluxa $$ for $ E < 2.3 \times 10^6 \GeV$, and $$ I^{\rm
bkgd}_{\nu_\mu+\bar\nu_\mu} \simle \left( 4.4 \, E_{\rm GeV}^{-3.69} + 3.9
\times 10^{-3} \, E_{\rm GeV}^{-3} \right) \persqcmssrGeV, \eqn \nufluxb $$ for
$ E > 2.3 \times 10^6 \GeV$.

Comparison of the decay neutrino flux~\unuflux\ with the atmospheric background
at $E=E_0$ gives the maximum explorable lifetime for  unclustered WUMPs  $$
\tau_x \simge \cases{  1.4 \times 10^{22} \sec \, { \displaystyle  b_\nu
m_{10}^{1.69} \over \displaystyle 1 + 2 \times 10^{-5} m_{10}^{1.04} }, & for $
m_x < 6.9 \times 10^{15} \eV$, \cr 1.4 \times 10^{22} \sec \, { \displaystyle
b_\nu  m_{10}^{1.69} \over \displaystyle 1 + 2 \times 10^{-3} m_{10}^{0.69} },
& for $ m_{10} > 0.6.9 \times 10^{15} \eV$. \cr} \eqn \unutau $$ This lifetime
is shown in Fig.~\figone\ as the heavy solid line labelled $\nu$, drawn for
$b_\nu=b_{\nu_\mu}+b_{\bar\nu_\mu}=2$.

For WUMPs clustered in our galactic halo, no integration over redshift needs to
be performed and the decay neutrino flux is given by $$  I_\nu(E) = b_\nu \, {
1 \over 4 \pi} \, { 1 \over m_x^2 \tau_x} \, \int { \rho_{\rm h}({\bf x}) \over
| {\bf x} - {\bf x}_\odot |^2 } d^3 x \, \delta(E-E_0) . \eqn \cnuflux $$
Assuming a detector with 10\% energy resolution to smear the delta function
the accessible lifetimes are a factor 15
larger than those for unclustered WUMPs. These lifetimes are shown in
Fig.~\figone, again for $b_\nu=2$, as the light solid line labelled $\nu$.

Present neutrino detectors are already able to exclude a sizeable region of the
WUMP mass-lifetime plane. The dotted line shows the lower bound on $\tau_x$
obtained in ref.~[\GGS] using IMB and Fly's Eye published data. Analogous
calculations show that a future 1 km$^2$ neutrino telescope would be able to
explore the region limited by the dotted line labelled $10^6$ at the flux level
of 1 muon per year.

\PLsection{3.~~Conclusions}

If the dark matter WUMPs are uniformly distributed in the universe, a
background-limited neutrino detector would be able to explore longer lifetimes
than those accessible to photon searches for any WUMP masses larger than
$ \sim
700 \GeV$.

If the WUMPs constitute the galactic halo, the longest lifetimes would be
probed by electron-positron searches for $m_x \simle 10 \TeV$. This occurs
because of the effective magnetic confinement of electrons and positrons in the
galaxy. Unfortunately, the containment time is quite poorly known and
the bounds so obtained would be subject to uncertainty. Searches for decay
neutrinos seem again the most powerful for WUMP masses above $\sim 700 \GeV$.

In both cases, a 1 km$^2$ neutrino telescope might be able to reach
lifetimes of order $10^{30} \sec$ for masses $m_x \simge 100 \TeV$.
Such a detector would be quite a good probe of
long-lived dark matter particles.

\PLsection{Acknowledgements}

I would like to thank Subir Sarkar for valuable discussions.

\vfill\eject

\refout
\vfill \eject

\figout

\bye